\documentclass[letter]{ieice}
\usepackage[fleqn]{amsmath}
\usepackage[varg]{txfonts}

\usepackage{graphicx}
\usepackage{graphics}
\usepackage{algorithm}
\usepackage{algorithmic}
\usepackage{multirow}
\usepackage{amsmath}
\usepackage{epstopdf}
\usepackage{graphicx}
\usepackage{color}
\usepackage{amssymb}

\DeclareMathOperator*{\diag}{diag}

\field{D}
\title{An Improved Traffic Matrix Decomposition Method with Frequency-Domain Regularization}
\titlenote{This work is supported by the National Natural Science Foundation of
China (Nos. 61073013 and 90818024)
and the Open Project of State Key Laboratory of Software Development Environment (No. SKLSDE-2012ZX-15).}
\authorlist{
 \authorentry[wangzhe@cse.buaa.edu.cn]{Zhe WANG}{n}{labelA}
 \authorentry{Kai HU}{m}{labelA}
 \authorentry{Baolin YIN}{n}{labelA}
}
\affiliate[labelA]{The authors are with the State Key Laboratory of Software Development Environment,
the School of Computer Science and Engineering,
Beihang University, Beijing, China.}

\received{2012}{9}{12}

\begin{document}
\maketitle
\begin{summary}
We propose a novel network traffic matrix decomposition method named
\emph{Stable Principal Component Pursuit with Frequency-Domain Regularization} (SPCP-FDR),
which improves the Stable Principal Component Pursuit (SPCP) method
by using a frequency-domain noise regularization function.
An experiment demonstrates the feasibility of this new decomposition method.
\end{summary}

\section{Introduction}
The network traffic matrix has been applied to many significant application problems such as capacity planning,
traffic engineering and anomaly detection.
A traffic matrix combines diverse traffic components with distinct temporal properties.
Therefore,
it is necessary to decompose them efficiently, and this problem is named traffic matrix structural analysis \cite{Lakhina1}.
We presented the traffic matrix decomposition model in \cite{Wang} and decomposed a traffic matrix into three sub-matrices,
which is equivalent to the generalized \emph{Robust Principal Component Analysis} (RPCA) problem \cite{Ma_SPCP}.
The results in \cite{Wang} were achieved by applying the \emph{Stable Principal Component Pursuit} (SPCP) method in \cite{Ma_SPCP}.

In this study,
we improve the traffic matrix decomposition method by using frequency-domain regularization.
This method is a variation of SPCP,
and is named \emph{Stable Principal Component Pursuit with Frequency-Domain Regularization} (SPCP-FDR).
We design the numerical algorithm for SPCP-FDR,
evaluate its decomposition results on the Abilene dataset \cite{Data1},
and show that SPCP-FDR achieves more rational traffic decompositions compared with SPCP.

\section{Background}
The standard RPCA problem is formally defined in \cite{Ma_RPCA1}:

\noindent\textbf{Problem 1} (Standard RPCA)
\emph{Suppose that a known matrix $X\in \mathbb{R}^{m\times n}$ is of the form $A + E$,
where $A$ and $E$ are unknown matrices.
It is assumed that $A$ has a low rank and $E$ is sparse.
The problem is to recover $A$ and $E$.}

In many applications,
data matrices satisfying the assumptions in Problem 1 are polluted by dense noise with small magnitude.
This leads to the generalized RPCA \cite{Ma_SPCP}:

\noindent\textbf{Problem 2} (Generalized RPCA)
\emph{Suppose that a known matrix $X\in \mathbb{R}^{m\times n}$ is of the form $A + E + N$,
where $A$, $E$ and $N$ are unknown matrices.
It is assumed that $A$ has a low-rank, $E$ is sparse,
and $N$ is an i.i.d. entry-wide noise matrix with small magnitude.
The problem is to recover $A$, $E$ and $N$.}

We refer RPCA as the generalized version as follows.
Zhou et al. \cite{Ma_SPCP} proved that under surprising board conditions,
for "almost all" data matrix $X$ which is the sum of a low-rank matrix and a sparse matrix,
and is corrupted by a dense noise matrix $N$ whose Frobenius norm $\|N\|_{F}\leq\delta$ ($\delta$ is a positive constant),
one could stably estimate $A$ and $E$ with high probability using this convex program:
\begin{equation}\label{SPCP_inequality}
\min_{A,E} \|A\|_{\ast} + \lambda\|E\|_{1} \ \ \ \ s. t. \|X - A + E\|_{F}\leq\delta,
\end{equation}
where $\lambda$ is a positive parameter,
$\|\cdot\|_{\ast}$ and $\|\cdot\|_{1}$ denote the matrix nuclear norm and the $l_{1}$ norm, respectively.
They named this method as \emph{Stable Principal Component Pursuit} (SPCP),
which needs to solve a time-consuming constrained program.
Thus they turned to solve another similar unconstrained program instead ($\tau$ is another positive parameter):
\begin{equation}\label{SPCP}
\min_{A,E} \tau(\|A\|_{\ast} + \lambda\|E\|_{1}) + \frac{1}{2}\|X-A-E\|_{F}^{2}.
\end{equation}

Suppose $X\in\mathbb{R}^{T\times P}$ is a traffic matrix,
and each column $X_{j}\in\mathbb{R}^{T}$ ($1\leq j\leq P$) is the traffic time-series of an Original-Destination (OD) flow.
Following the traffic matrix decomposition model \cite{Wang},
$X$ is the sum of three sub-matrices:
(1) The deterministic traffic matrix $A$ is a low-rank matrix,
contributed by the periodic traffic changes in each OD flow;
(2) The anomaly traffic matrix $E$ is a sparse matrix,
but its nonzero entries may have large magnitudes;
(3) The noise matrix $N$ is constituted of independent random variables with relatively small magnitudes,
the entries in one column constitute a white noise vector,
but those in different columns have distinct variances.
In \cite{Wang}, we estimated the noise traffic variances $\{\sigma_{j}\}_{j=1}^{P}$ of all the OD flows in $X$,
and divided $X_{j}$ by $\sigma_{j}$ ($1\leq j\leq P$).
This preprocessing normalizes random variables in $N$,
and preserves the rank of $A$,
as well as the sparsity of $E$.
Thus traffic matrix decomposition is equivalent to RPCA,
and we only consider traffic matrices with unit noise traffic variance.

\section{Stable Principal Component Pursuit with Frequency-Domain Regularization}
Suppose "$\circledast$" is a traffic matrix decomposition method,
and $N^{\circledast}\in\mathbb{R}^{T\times P}$ is the decomposed noise traffic matrix of $X$ by $\circledast$.
In this letter, SPCP and SPCP-FDR are denoted as "$\odot$" and "$\oplus$", respectively.
We consider the frequency-domain property of the decomposed noise traffic matrices.
For each noise traffic time-series $N^{\circledast}_{j}$ (the $j-$th column of $N^{\circledast}$),
$\alpha_{j}\in\mathbb{C}^{T}$ denotes its Discrete Fourier Transform (DFT): 
\begin{equation}\label{FFT}
\alpha_{j} = W^{\textrm{T}}N^{\circledast}_{j} = \left[W_{1} \ W_{2} \ \cdots \ W_{T}\right]^{\textrm{T}} N^{\circledast}_{j},
\end{equation}
where $W\in\mathbb{C}^{T\times T}$ is the discrete Fourier basis matrix.
The $t-$th column vector $W_{t}\in\mathbb{C}^{T}$ ($1\leq t\leq T$) is defined as:
\begin{equation}\label{Fourier Basis}
W_{t}(k)=\frac{1}{\sqrt{T}}e^{-\frac{2\pi i}{T}(t-1)(k-1)} \ \ \ \ k = 1,2,...,T.
\end{equation}
$N^{\circledast}_{j}$'s spectral density $\varphi_{j}\in\mathbb{R}^{T}$ is defined as:
\begin{equation}\label{power_spectrum}
\varphi_{j}(t)=\left|\alpha_{j}(t)\right|^{2} \ \ \ \ t=1,2,...,T.
\end{equation}
As $N^{\circledast}_{j}$ is a real signal,
$\varphi_{j}(t)=\varphi_{j}(T-t+2)$ for each position $t\in[2,T]\cap \mathbb{N}$,
and $(T-t+2)$ is named the dual position of $t$.
The spectra $\varphi_{j}(t)$ at positions close to $2$ or $T$ describe $N^{\circledast}_{j}$'s
power distributed in low-frequency domain;
conversely, the spectra at positions close to $(\frac{T}{2}+1)$ indicate the high-frequency power.
Furthermore, the spectral density $\Phi_{N^{\circledast}}\in\mathbb{R}^{T}$ of $N^{\circledast}$
is defined as the entry-wise sum of all the noise traffic time-series' spectral density:
\begin{equation}\label{power_spectrum_N}
\Phi_{N^{\circledast}}(t) = \sum_{j=1}^{P}\varphi_{j}(t) \ \ \ \ t = 1,2,...,T.
\end{equation}

\begin{figure}[ht]
\centering
\scalebox{0.42}[0.42]{\includegraphics*[15,303][580,515]{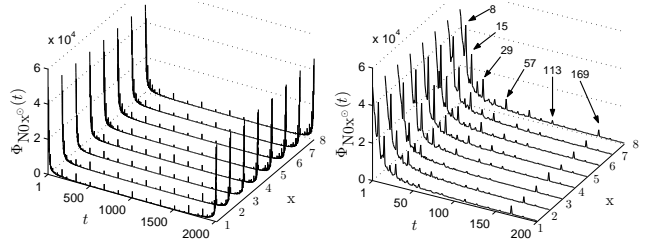}}
\vspace{-5pt}
\begin{center}
\caption{The spectral density of $\textrm{N}01^{\odot}\sim\textrm{N}08^{\odot}$ (left);
For each noise traffic matrix,
the first 200 positions of its spectral density are specially magnified (right),
which describe the low-frequency power.}\label{power_spectrum_spcp}
\end{center}
\vspace{-10pt}
\end{figure}

We compute the spectral density of eight noise traffic matrices $\textrm{N}01^{\odot}\sim\textrm{N}08^{\odot}$
and display them in Fig. \ref{power_spectrum_spcp},
which are independently decomposed from the eight Abilene \cite{Data1} weekly traffic matrices
$\textrm{X}01\sim\textrm{X}08$ by SPCP we adopted in \cite{Wang}.
In this dataset, $P=121$, $T=2016$, and the minimal time interval is 5 minutes.
Thus for each noise traffic matrix $\textrm{N0x}^{\odot}$ ($\textrm{x}\in\{1,2,...,8\}$),
$\Phi_{\textrm{N0x}^{\odot}}(t)$ represents the power captured by periodic traffic with period $\frac{2016}{\min(t-1,2016-t+2)}\frac{5}{60}$ hour(s).
We discover these two properties for the noise traffic matrices decomposed by SPCP:
(1) The low-frequency spectra are generally much larger than the high-frequency spectra.
This means that most energy of these noise traffic matrices is contributed by the low-frequency traffic patterns;
(2) Quite a few low-frequency spectra have dramatically larger values than their neighbors
(No. 8, 15, 29, 57, 113 and 169 spectra, as well as the spectra at dual positions,
and their corresponding periods are 24, 12, 6, 3, 1.5 and 1 hour(s), respectively).

These properties match our assumptions of the noise traffic poorly.
As each column of the noise traffic matrix is assumed as a white noise vector,
if the noise traffic is exactly recovered by SPCP,
the spectral density should show a flat distribution.
We briefly explain the cause of this limitation.
In most cases,
the volume of a traffic matrix is mainly contributed by the deterministic traffic.
The deterministic traffic changes slowly and shows typical diurnal pattern.
Therefore, the spectral density of each OD flow presents unbalanced distribution:
The low-frequency spectra usually have much larger values,
and the spectra whose periods are the factors of 24 hours show strong peaks.
The optimization problem (\ref{SPCP}) can be written in this equivalent form:
\begin{equation}\label{SPCP_constraint}
\min_{A, E, N} \|A\|_{\ast} + \lambda\|E\|_{1} + \gamma\|N\|_{F}^{2} \ \ s.t. A+E+N=X,
\end{equation}
where $\gamma=1/2\tau$,
and the objective function for $N$ is the Frobenius norm.
Thus problem (\ref{SPCP_constraint}) has no frequency-domain consideration on the noise traffic.
With high probability,
the noise traffic decomposed by SPCP inherits the unbalanced spectral density distribution of the OD flow traffic.

These discussions motivate us to replace $\|N\|_{F}^{2}$ by a new objective function which focuses on $N$'s frequency-domain property,
and the new decomposed noise traffic should have more flat spectral density.
We define the frequency-domain weight matrix $C = \diag(c_{1},...,c_{T})$ satisfying:
\begin{equation}\label{weight_assumption}
c_{t}>0,\ t=1,2,...,T; \ \ \sum_{t=1}^{T}c_{t}^{2}=T.
\end{equation}
We propose the new traffic decomposition method called
\emph{Stable Principal Component Pursuit with Frequency-Domain Regularization} (SPCP-FDR).
Compared with SPCP,
SPCP-FDR adopts the same objective functions for the deterministic and the anomaly traffic, respectively;
while $\|CW^{\textrm{T}}N\|_{F}$ is chosen as the new objective function for the noise traffic
($C$'s design is presented in Section 4):
\begin{equation}\label{SPCP-FDR}
\min_{A, E, N} \|A\|_{\ast} + \lambda\|E\|_{1} + \gamma\|CW^{\textrm{T}}N\|_{F}^{2} \ \ s.t. A+E+N=X,
\end{equation}
where $\lambda$ and $\gamma$ are positive parameters balancing the three objective functions.
We choose $\lambda = 1/\sqrt{\max(T, P)}$ because it is demonstrated as a proper choice in \cite{Ma_RPCA1}.
For the choice of $\gamma$, consider the simplest case:
each column in $N$ is a standard Gaussian white noise.
The DFT of one Gaussian white noise is also a Gaussian white noise,
and it can be proved that $\mathbb{E}\left[\|CW^{\textrm{T}}N\|_{F}^{2}\right]=\mathbb{E}\left[\|N\|_{F}^{2}\right]$
($\mathbb{E}$ denotes expectation).
Therefore, SPCP-FDR can be seen as an approximation of SPCP under the Gaussian white noise assumption.
In \cite{Wang}, we choose $\tau=\sqrt{2\log(TP)\max(T,P)}$ for problem (\ref{SPCP}).
Thus in this study $\gamma=1/2\tau=1/\left(2\sqrt{2\log(TP)\max(T,P)}\right)$.

\section{Implementation Details}
\subsection{The Design of Frequency-Domain Weight Matrix}
We design $C$ for decomposing the Abilene weekly traffic matrices ($T=2016$),
and we believe that the key technologies are adaptable to other datasets with small modifications.
Instinctively, it punishes the low-frequency spectra,
especially for the spectra whose corresponding periods are the factors of 24 hours.
Define the position set $S_{1}$ as:
\begin{equation*}
\begin{split}
& S_{1}=S_{1}^{a}\cup S_{1}^{b};\\
& S_{1}^{a}=\{8, 15, 29, 57, 113, 169\}; \ S_{1}^{b}=\{t \ | \ (T-t+2)\in S_{1}^{a}\}.
\end{split}
\end{equation*}
$S_{1}^{b}$ represents the dual positions of $S_{1}^{a}$.
We design $\{c_{t}\}_{t=1}^{T}$ as:
\begin{equation}
c_{t}=
\begin{cases}
\beta[v(t)+\rho] & t\in S_{1} \\
\beta v(t)       & \textrm{otherwise}
\end{cases},
\end{equation}
where $v(x)$ is a positive function in $[1,T]$ satisfying $v(x)=v(T-x+2)$ in $[2,T]$.
Meanwhile, it decreases monotonically in $[1,\frac{T}{2}+1]$.
$\rho>0$ is an additional penal parameter for the spectra correlated to $S_{1}$.
$\beta>0$ is a scaling parameter.
In this study, we choose
\begin{equation}
v(x)=
\begin{cases}
4e^{-\frac{(x-1)}{200}}+1    & x\in [1,\frac{T}{2}+1) \\
4e^{-\frac{(T-x+1)}{200}}+1  & x\in [\frac{T}{2}+1,T]
\end{cases}
\end{equation}
and $\rho=2$.
At last, we get $\beta=0.4832$ from assumption (\ref{weight_assumption}) for this special choice of $v(x)$ and $\rho$.

\subsection{The APG Algorithm for SPCP-FDR}
The \emph{Accelerated Proximal Gradient} (APG) algorithm for SPCP-FDR solves a relaxed approximation problem of (\ref{SPCP-FDR}):
\begin{equation}\label{APG}
\begin{split}
  & \min_{A,E,N}F(A, E, N)=\min_{A,E,N}\mu g(A,E,N)+f(A,E,N)\\
  & g(A,E,N)\triangleq\|A\|_{\ast} + \lambda\|E\|_{1} + \gamma\|CW^{\textrm{T}}N\|_{F}^{2},\\
  & f(A,E,N)\triangleq\frac{1}{2}\|A+E+N-X\|_{F}^{2},
\end{split}
\end{equation}
where $\mu>0$ is a parameter.
As $\mu\rightarrow0$, the solution to (\ref{APG}) approaches to the solution to (\ref{SPCP-FDR}).
Not directly minimizing $F(A,E,N)$,
this algorithm minimizes a sequence of quadratic approximations $Q(A,E,N,Y^{A},Y^{E},Y^{N})$ to $F(A,E,N)$
at point $(Y^{A}, Y^{E}, Y^{N})$ (renewed in each step):\\
\begin{equation}\label{APG_quadratic_approximations}
\begin{split}
           & Q(A,E,N,Y^{A},Y^{E},Y^{N}) \\
\triangleq & \mu g(A,E,N) + f(Y^{A},Y^{E},Y^{N}) + \\
           & \langle\nabla f(Y^{A},Y^{E},Y^{N}),(A,E,N)-(Y^{A},Y^{E},Y^{N})\rangle + \\
           & \frac{L_{f}}{2}\|(A,E,N)-(Y^{A},Y^{E},Y^{N})\|_{F}^{2},
\end{split}
\end{equation}
where the Lipschitz constant $L_{f}=3$.
It can be derived that:\\
\begin{equation}
\begin{split}\label{APG_subproblem}
 &\min_{A,E,N} Q(A,E,N,Y^{A},Y^{E},Y^{N}) \\
=&\min_{A}\frac{L_{f}}{2}\|A-G^{A}\|_{F}^{2}+\mu\|A\|_{\ast}+ \\
 &\min_{E}\frac{L_{f}}{2}\|E-G^{E}\|_{F}^{2}+\mu\lambda\|E\|_{1}+ \\
 &\min_{N}\frac{L_{f}}{2}\|N-G^{N}\|_{F}^{2} + \mu\gamma\|CW^{\textrm{T}}N\|_{F}^{2} + \textrm{constant},
\end{split}
\end{equation}\\
where
\begin{equation}
G^{\square} = Y^{\square}-\frac{1}{L_{f}}(Y^{A}+Y^{E}+Y^{N}-X), \ \ \square\in\{A, E, N\}.
\end{equation}
Problem (\ref{APG_subproblem}) splits into three independent optimization problems,
and the first two (for $A$ and $E$) are well studied \cite{Ma_APG}.
We give the solution to the third problem by derivation:
\begin{equation}\label{N}
N = L_{f}\left[L_{f}I_{T\times T}+2\mu\gamma WC^{2}W^{\textrm{T}}\right]^{-1}G^{N}.
\end{equation}

Following the main idea in \cite{Ma_APG},
we present the APG algorithm for the SPCP-FDR method in Algorithm \ref{algo1}.
This algorithm has the $O(1/k^{2})$ convergence rate.
Because the proof is very close to Theorem 2.1 in \cite{Ma_APG} and Theorem 4.4 in \cite{Beck_FISTA},
we omit it and directly summarize this result:

\noindent\textbf{Theorem 1} Let $F(A,E,N)=\overline{\mu}g(A,E,N)+f(A,E,N)$.
Then for all $k>k_{0}\triangleq\left\lceil\log\left(\frac{\mu_{0}}{\overline{\mu}}\right)/\log\left(\frac{1}{\eta}\right)\right\rceil$,
we have
\begin{equation}
F(X_{k})-F(X^{\ast})\leq 6\|X_{k_{0}}-X^{\ast}\|_{F}^{2}/(k-k_{0}+1)^{2},
\end{equation}
where $X_{k}=(A_{k},E_{k},N_{k})$ is defined in Algorithm \ref{algo1},
and $X^{\ast}=(A^{\ast},E^{\ast},N^{\ast})$ is a solution to (\ref{APG}) when $\mu=\overline{\mu}$.
Notice that in this study $L_{f}=3$, while in \cite{Ma_APG} it equals to 2.

\vspace{-10pt}
\begin{algorithm}[ht]
\caption{\small The APG Algorithm for SPCP-FDR}\label{algo1}
\textbf{Input}: traffic matrix $X\in\mathbb{R}^{T\times P}$ with unit noise variance.

\textbf{Initialization:} $A_{0}=A_{-1}=E_{0}=E_{-1}=N_{0}=N_{-1}=\textbf{0}^{T\times P}$; $t_{0}=t_{-1}=1$;
$\mu_{0}=0.99\|X\|_{2}$; $\overline{\mu}=10^{-5}\mu_{0}$; $k=0$.


$\eta=0.9$; $L_{f}=3$; $\lambda = 1/\sqrt{\max(T, P)}$; $\gamma=1/\left(2\sqrt{2\log(TP)\max(T,P)}\right)$.

\textbf{While} \ not converged \textbf{do}

\quad $Y_{k}^{A}=A_{k}+\frac{t_{k-1}-1}{t_{k}}(A_{k}-A_{k-1})$; $Y_{k}^{E}=E_{k}+\frac{t_{k-1}-1}{t_{k}}(E_{k}-E_{k-1})$;

\quad $Y_{k}^{N}=N_{k}+\frac{t_{k-1}-1}{t_{k}}(N_{k}-N_{k-1})$;

\quad $G^{\square}_{k} = Y^{\square}_{k}-\frac{1}{L_{f}}(Y^{A}_{k}+Y^{E}_{k}+Y^{N}_{k}-X)$, $\square\in\{A, E, N\}$;

\quad $(U,S,V)=\mathrm{svd}\left[G_{k}^{A}\right]$; \ //$\mathrm{svd}[\cdot]$ denotes singular value decomposition.

\quad $A_{k+1}=U\mathcal{S}_{\frac{\mu_{k}}{L_{f}}}\left[S\right]V^{\textrm{T}}$;
      $E_{k+1}=\mathcal{S}_{\frac{\lambda\mu_{k}}{L_{f}}}\left[G_{k}^{E}\right]$;

\quad //$\mathcal{S}_{\epsilon}[\cdot]$ denotes soft-thresholding operator with threshold $\epsilon>0$.

\quad $N_{k+1} = L_{f}\left[L_{f}I_{T\times T}+2\mu_{k}\gamma WC^{2}W^{\textrm{T}}\right]^{-1}G_{k}^{N}$;

\quad $t_{k+1}=(1+\sqrt{4t_{k}^{2}+1})/2$; $\mu_{k+1}=\max(\eta\mu_{k},\overline{\mu})$; $k=k+1$.

\textbf{End while}

\textbf{Output}: $A=A_{k}$; $E=E_{k}$; $N=E_{k}$.
\end{algorithm}
\vspace{-20pt}

\section{Experiment Results}
\begin{figure}[ht]
\centering
\scalebox{0.44}[0.44]{\includegraphics*[15,298][573,525]{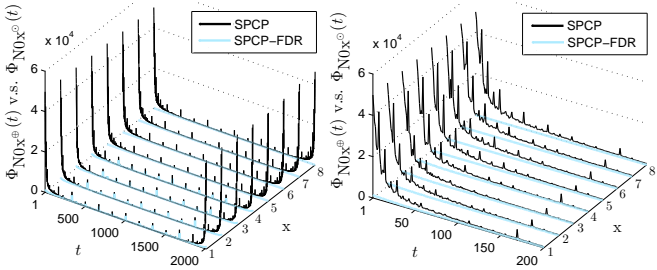}}
\vspace{-5pt}
\begin{center}
\caption{The spectral density of $\textrm{N}01^{\oplus}\sim\textrm{N}08^{\oplus}$ decomposed by SPCP-FDR,
compared with the spectral density of $\textrm{N}01^{\odot}\sim\textrm{N}08^{\odot}$ decomposed by SPCP (left);
For each matrix,
the first 200 positions of its spectral density are specially magnified (right),
which describe the low-frequency power.}\label{power_spectrum_spcp_fdr}
\end{center}
\vspace{-10pt}
\end{figure}

\begin{table}[htb]
\caption{A comparison between SPCP and SPCP-FDR on the decomposed deterministic traffic matrices.}\label{deterministic_traffic}
\begin{center}
\begin{tabular}{lllll}
\hline
Traffic matrix  & rank$(\textrm{X}0\textrm{x})$ & rank$(\textrm{A}0\textrm{x}^{\odot})$ & rank$(\textrm{A}0\textrm{x}^{\oplus})$ & $\frac{\|\textrm{A}0\textrm{x}^{\oplus}\|_{F}}{\|\textrm{A}0\textrm{x}^{\odot}\|_{F}}$\\
\hline
$\textrm{X01} \ (\textrm{x}=1)$  &121 &10 & 19 & 1.0158\\
$\textrm{X02} \ (\textrm{x}=2)$  &121 &11 & 22 & 1.0171\\
$\textrm{X03} \ (\textrm{x}=3)$  &121 &12 & 20 & 1.0283\\
$\textrm{X04} \ (\textrm{x}=4)$  &121 &11 & 18 & 1.0145\\
$\textrm{X05} \ (\textrm{x}=5)$  &121 &10 & 19 & 1.0148\\
$\textrm{X06} \ (\textrm{x}=6)$  &121 &10 & 22 & 1.0162\\
$\textrm{X07} \ (\textrm{x}=7)$  &121 &13 & 24 & 1.0210\\
$\textrm{X08} \ (\textrm{x}=8)$  &121 &12 & 23 & 1.0182\\
\hline
\end{tabular}
\end{center}
\vspace{-10pt}
\end{table}
We evaluate the SPCP-FDR method by using the Abilene traffic matrices $\textrm{X}01\sim\textrm{X}08$.
For each $\textrm{x}\in\{1,...,8\}$, suppose $\textrm{X}0\textrm{x}$ is decomposed as
$\{\textrm{A}0\textrm{x}^{\oplus}, \textrm{E}0\textrm{x}^{\oplus}, \textrm{N}0\textrm{x}^{\oplus}\}$
and $\{\textrm{A}0\textrm{x}^{\odot}, \textrm{E}0\textrm{x}^{\odot}, \textrm{N}0\textrm{x}^{\odot}\}$
by SPCP-FDR and SPCP, respectively.

Figure \ref{power_spectrum_spcp_fdr} compares the spectral density of the noise traffic matrices
$\textrm{N}01^{\oplus}\sim\textrm{N}08^{\oplus}$ (blue) and $\textrm{N}01^{\odot}\sim\textrm{N}08^{\odot}$ (black).
For each $\textrm{x}\in\{1,...,8\}$,
$\textrm{N0x}^{\oplus}$ has more flat spectral density distribution than $\textrm{N0x}^{\odot}$.
As we further regularize noise traffic's spectra whose positions lie in $S_{1}$,
their magnitudes dramatically decline.
Therefore, for the new decompositions of $\textrm{X}01\sim\textrm{X}08$ by SPCP-FDR,
most low-frequency traffic pattern is efficiently eliminated from the resulting noise traffic matrices $\textrm{N}01^{\oplus}\sim\textrm{N}08^{\oplus}$.
Table \ref{deterministic_traffic} compares the deterministic traffic matrices
$\textrm{A}01^{\oplus}\sim\textrm{A}08^{\oplus}$ and $\textrm{A}01^{\odot}\sim\textrm{A}08^{\odot}$.
The ranks of $\textrm{A}01^{\oplus}\sim\textrm{A}08^{\oplus}$ decomposed by SPCP-FDR are nearly twice as those decomposed by SPCP.
However, as all these values do not exceed 24 and the original traffic matrices' ranks are 121,
$\textrm{A}01^{\oplus}\sim\textrm{A}08^{\oplus}$ still satisfy the low-rank characteristic.
For each $\textrm{x}\in\{1,...,8\}$,
the Frobenius norm of $\textrm{A}0\textrm{x}^{\oplus}$ is a little larger than that of $\textrm{A}0\textrm{x}^{\odot}$,
and it can be explained that most low-frequency traffic pattern eliminated from the noise traffic is added to the deterministic traffic.


\begin{figure}[ht]
\centering
\scalebox{0.5}[0.5]{\includegraphics*[60,367][530,475]{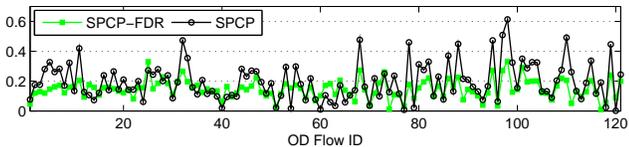}}
\vspace{-5pt}
\begin{center}
\caption{A comparison between SPCP and SPCP-FDR on the absolute Pearson correlation coefficients
(between the deterministic traffic time-series and the noise traffic time-series of each OD flow)
of $\textrm{X}01$.}\label{cross_correlation}
\end{center}
\vspace{-10pt}
\end{figure}

\begin{figure}[ht]
\centering
\scalebox{0.43}[0.43]{\includegraphics*[30,296][572,538]{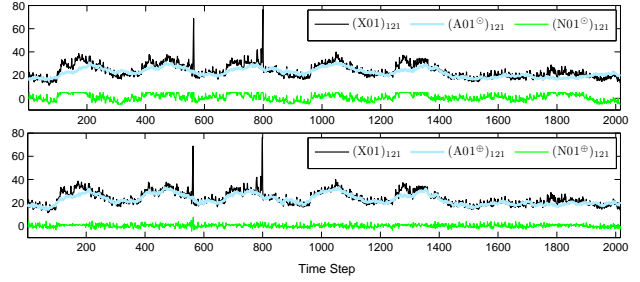}}
\vspace{-5pt}
\begin{center}
\caption{A comparison between SPCP and SPCP-FDR on the decomposition result of OD flow $(\textrm{X}01)_{121}$.
Upper panel: the deterministic and the noise traffic time-series decomposed by SPCP;
Bottom panel: the deterministic and the noise traffic time-series decomposed by SPCP-FDR.}
\label{X_{121}_decomposition}
\end{center}
\vspace{-10pt}
\end{figure}
As a case study, we analyze the decomposition result of $\textrm{X}01$ in detail.
For each OD flow $(\textrm{X}01)_{j}$ ($1\leq j\leq 121$),
compute the absolute Pearson correlation coefficient between $(\textrm{A}01^{\oplus})_{j}$ and $(\textrm{N}01^{\oplus})_{j}$,
and the coefficient between $(\textrm{A}01^{\odot})_{j}$ and $(\textrm{N}01^{\odot})_{j}$.
Figure \ref{cross_correlation} displays these coefficients arranged by flow ID.
This is a proper metric of the cross-correlation and it should be as small as possible.
For most OD flows in $\textrm{X}01$, compared with SPCP,
SPCP-FDR's absolute Pearson correlation coefficients show significant decline.
Actually, $\textrm{X}02\sim\textrm{X}08$ show quite similar results.
Thus SPCP-FDR could efficiently reduce this cross-correlation.
Figure \ref{X_{121}_decomposition} compares SPCP-FDR with SPCP on decomposition result of the No. 121 OD flow,
which produces the largest volume in $\textrm{X}01$.
Instinctively, the noise traffic time-series $(\textrm{N}01^{\odot})_{121}$ contains a distinct diurnal trend,
which should be classified as the deterministic traffic better.
By contrast, $(\textrm{N}01^{\oplus})_{121}$ presents a more stable temporal pattern.
Thus SPCP-FDR gives a more rational decomposition for this flow.

\section{Conclusions}
This letter presents a novel traffic matrix decomposition method named SPCP-FDR,
which improves SPCP by using frequency-domain regularization.
We propose an APG algorithm for SPCP-FDR, its convergence rate is $O(1/k^{2})$, and demonstrate it on a real-world dataset.
SPCP-FDR efficiently eliminates the low-frequency periodic traffic from the noise traffic,
maintains deterministic traffic's low-rank property,
and shows lower cross-correlation between these two kinds of traffic than SPCP.
Therefore, SPCP-FDR achieves more rational traffic decompositions.


\begin{thebibliography}{99}
\bibitem{Lakhina1}
A. Lakhina, K. Papagiannaki, M. Crovella, C. Diot, E. D. Kolaczyk, and N. Taft.
Structural analysis of network traffic flows.
SIGMETRICS Perform. Eval. Rev. vol. 32, no. 1, pp. 61-72, June 2004.

\bibitem{Wang}
Z. Wang, K. Hu, K. Xu, B. Yin and X. Dong.
Structural Analysis of Network Traffic Matrix via Relaxed Principal Component Pursuit.
Computer Networks, vol. 56, no. 7, pp. 2049-2067, 2012.

\bibitem{Ma_SPCP}
Z. Zhou, X. Li, J. Wright, E. Candes, and Y. Ma.
Stable principal component pursuit.
In Proc. IEEE ISIT 2010.

\bibitem{Ma_RPCA1}
E. Candes, X. Li, Y. Ma, and J. Wright. Robust principal component analysis?
Journal of the ACM. vol. 58, no. 3, pp. 1-37, 2011.

\bibitem{Ma_APG}
Z. Lin, A. Ganesh, J. Wright, L. Wu, M. Chen and Y. Ma.
Fast convex optimization algorithms for exact recovery of a corrupted low-rank matrix.
In Proc. IEEE CAMSAP 2009.

\bibitem{Beck_FISTA}
A. Beck and M. Teboulle.
A fast iterative shrinkage-thresholding algorithm for linear inverse problems.
SIAM Journal on Imaging Sciences, vol. 2, no. 1, pp. 183-202, 2009.

\bibitem{Data1}
Abilene dataset: http://www.cs.utexas.edu/$^{\sim}$yzhang/research/AbileneTM/
\end{thebibliography}

\profile{}{}

\end{document}